\documentclass[a4paper,fleqn,usenatbib]{mnras}

\usepackage{newtxtext,newtxmath}

\usepackage[T1]{fontenc}
\usepackage{ae,aecompl}

\usepackage{graphicx}	
\usepackage{amsmath}	
\usepackage{amssymb}	

\title[Does the $L_{\rm MIR}$--$L_{\rm X}$ relation for AGNs depend on $\lambda_{\rm Edd}$?]{Does the mid-infrared--hard X-ray luminosity relation for active galactic nuclei depend on Eddington ratio?}

\author[Y. Toba et al.]{
Yoshiki Toba,$^{1,2,3}$\thanks{E-mail: toba@kusastro.kyoto-u.ac.jp}
Yoshihiro Ueda,$^{1}$
Kenta Matsuoka,$^{3,4,5}$
Megumi Shidatsu,$^{6}$
\newauthor
Tohru Nagao,$^{3}$
Yuichi Terashima,$^{6}$
Wei-Hao Wang,$^{2}$
and Yu-Yen Chang$^{2}$
\\
$^{1}$Department of Astronomy, Kyoto University, Kitashirakawa-Oiwake-cho, Sakyo-ku, Kyoto 606-8502, Japan\\
$^{2}$Academia Sinica Institute of Astronomy and Astrophysics,\\ 11F of Astronomy-Mathematics Building, AS/NTU, No.1, Section 4, Roosevelt Road, Taipei 10617, Taiwan\\
$^{3}$Research Center for Space and Cosmic Evolution, Ehime University, 2-5 Bunkyo-cho, Matsuyama, Ehime 790-8577, Japan\\
$^{4}$Dipartimento di Fisica e Astronomia, Universit\'a degli Studi di Firenze, via G. Sansone 1, 50019 Sesto Fiorentino, Italy\\
$^{5}$INAF-Osservatorio Astrofisico di Arcetri, Largo Enrico Fermi 5, 50125 Firenze, Italy \\
$^{6}$Department of Physics, Ehime University, 2-5 Bunkyo-cho, Matsuyama, Ehime 790-8577, Japan
}

\date{\today}

\pubyear{2018}

\begin{document}
\label{firstpage}
\pagerange{\pageref{firstpage}--\pageref{lastpage}}
\maketitle

\begin{abstract}
We revisit the correlation between the mid-infrared (6 $\micron$) and hard X-ray (2--10 keV) luminosities of active galactic nuclei (AGNs) to understand the physics behind it.
We construct an X-ray flux-limited sample of 571 type 1 AGNs with $f_{0.5-2.0 \,{\rm keV}} > 2.4 \times 10^{-12}$ erg cm$^{-2}$ s$^{-1}$, drawn from the {\it ROSAT} Bright Survey catalog.
Cross-matching the sample with infrared data taken from {\it Wide-field Infrared Survey Explorer}, we investigate the relation between the rest-frame 6 $\micron$ luminosity ($L_{\rm 6}$) and the rest-frame 2--10 keV luminosity ($L_{\rm X}$),  where $L_{\rm 6}$ is corrected for the contamination of host galaxies by using the spectral energy distribution fitting technique.
We confirm that $L_{\rm 6}$ and $L_{\rm X}$ are correlated over four orders of magnitude, in the range of $L_{\rm X} = 10^{42-46}$ erg s$^{-1}$.
We investigate what kinds of physical parameters regulate this correlation.
We find that $L_{\rm X}$/$L_{\rm 6}$ clearly depends on the Eddington ratio ($\lambda_{\rm Edd}$) as $\log \lambda_{\rm Edd} = -(0.56 \pm 0.10) \log \, (L_{\rm X}/L_{\rm 6}) - (1.07 \pm 0.05)$, even taking into account quasars that are undetected by {\it ROSAT} as well as those detected by {\it XMM-Newton} in the literature.
We also add hyper-luminous quasars with $L_{\rm 6}$ $>$ 10$^{46}$ erg s$^{-1}$ in the literature and perform a correlation analysis.
The resultant correlation coefficient is $-0.41 \pm 0.07$, indicating a moderately tight correlation between $L_{\rm X}$/$L_{\rm 6}$ and $\lambda_{\rm Edd}$.
This means that AGNs with high Eddington ratios tend to have lower X-ray luminosities with respect to the mid-infrared luminosities.
This dependence can be interpreted as a change in the structure of the accretion flow.
\end{abstract}

\begin{keywords}
infrared: galaxies  --- X-rays: galaxies --- methods: observational  --- methods: statistical --- quasars: general
\end{keywords}



\section{Introduction} 
\label{intro}

It is well-known that active galactic nuclei (AGNs) show a positive ``linear'' correlation between the mid-infrared (IR) and the hard X-ray luminosity in a log-log space \citep[e.g.,][]{Lutz,Horst,Fiore,Gandhi,Ichikawa_12,Matsuta,Asmus,Mateos,Isobe,Garca-Bernete,Ichikawa_17}.
In the context of the AGN unified model \citep[e.g.,][]{Antonucci,Urry}, the X-ray emission in an AGN is generated in a hot corona by inverse Compton scattering of thermal photons from an accretion disk \citep[e.g.,][]{Sunyaev,Haardt_91,Haardt_93}.
The disk emission, mainly in the ultraviolet (UV) and optical bands, is partly absorbed by dusty obscuring material, 
so-called the dust torus \citep{Krolik} outside the accretion disk, and the dust torus produces the reprocessed radiation in the near-IR (NIR) to mid-IR (MIR) band \citep[see ][and references therein]{Elitzur}.
AGNs show a tight correlation between their MIR and hard X-ray luminosities, and this correlation seems to be valid for both type 1 (unobscured) and type 2 (obscured) AGNs.  
Because the hard X-ray luminosity is considered to trace the intrinsic power of AGNs, the correlation suggests that the MIR luminosity is also a good tracer of the intrinsic AGN activity, regardless of the dust obscuration, and that the dust distribution of an AGN torus is clumpy rather than smooth \citep[see ][and references therein]{Horst,Gandhi}.

Recently, using AGNs spanning from low-luminosity Seyfert galaxies to luminous quasars, which are selected from the Sloan Digital Sky Survey \citep[SDSS:][]{York}, \cite{Stern} revisited the relationship between the rest-frame 6 \micron \, luminosity ($L_{\rm 6}$) and the absorption-corrected, rest-frame 2--10 keV luminosity ($L_{\rm X}$), where 
$L_{\rm 6}$ was estimated by using data taken from {\it Wide-field Infrared Survey Explorer} \citep[{\it WISE}:][]{Wright} while $L_{\rm X}$ was taken from \cite{Just}.
\cite{Stern} reported that the $L_{\rm 6}$--$L_{\rm X}$ correlation is ``non-linear'' in a log-log space, and is flattened in the high luminosity region.
Above a 6 \micron \, luminosity of $\log\, \nu L_{\nu}$ (6 \micron) $=$ 45 (where $\nu L_{\nu}$ (6 \micron) is in units of erg s$^{-1}$), the 2--10 keV luminosity was found to be lower than what was expected from the extrapolation of the 
$L_{\rm 6}$--$L_{\rm X}$ correlation determined in the lower $\nu L_{\nu}$ (6 \micron) region 
\citep[see also][]{Lanzuisi_09,Chen}.
Subsequent works focusing on high-luminosity quasars, such as hot dust-obscured galaxies \citep[hot DOGs:][]{Eisenhardt,Wu}, extremely red quasars \citep[ERQs:][]{Ross,Hamann}, and hyper-luminous quasars, selected from the SDSS and {\it WISE} \citep[WISSH quasars:][]{Bischetti,Duras}, supported this deficit of the X-ray luminosity with respect to the 6 \micron \, luminosity \citep[e.g.,][]{Martocchia,Ricci,Vito,Goulding}.

The origin of this deficit is still unclear, although a few possibilities, such as the luminosity dependence of the optical to X-ray flux ratio ($\alpha_{\rm OX}$), the 2--10 keV bolometric correction ($\kappa_{\rm 2-10 \,keV}$), and the Eddington ratio ($\lambda_{\rm Edd}$), have been suggested from a qualitative perspective (see Section \ref{Edd_D}).
Interpreting the deficit becomes even more complicated due to uncertainties in the absorption correction 
for the X-ray luminosities of obscured AGNs \citep[see e.g.,][]{Goulding}. 
In order to investigate the origin of the X-ray deficit and the physics behind the $L_{\rm 6}$--$L_{\rm X}$ correlation, more quantitative approaches are required, by using unobscured AGNs, whose intrinsic X-ray luminosities are less affected by absorption.

In this paper, we revisit the $L_{\rm 6}$--$L_{\rm X}$ correlation of X-ray selected type 1 (unobscured) AGNs, particularly focusing on $\lambda_{\rm Edd}$, and determine whether or not the $L_{\rm 6}$--$L_{\rm X}$ correlation depends on $\lambda_{\rm Edd}$.
Throughout this paper, the adopted cosmology is a flat universe with $H_0 =$ 70 km s$^{-1}$ Mpc$^{-1}$, $\Omega_{\rm M} =$ 0.3, and $\Omega_{\Lambda} =$ 0.7.
Unless otherwise noted, we use $L_{6}$ and $L_{\rm X}$ as a shorthand alias for $\nu L_{\nu}$ (6 \micron) and $L$ (2--10 keV), respectively, and the luminosity at any wavelength/energy-band is given in units of erg s$^{-1}$.

\section{Data and analysis} 
\label{DA}

\subsection{Sample selection}
We sampled X-ray selected type 1 AGNs from the {\it ROSAT} Bright Survey (RBS) catalog \citep{Fischer,Schwope}, which provided an identification of the 2072 X-ray sources in 0.5--2.0 keV detected at $|b|$ $>$ 30$\degr$ during the ROSAT All-Sky Survey (RASS).
The flux limit of RBS is $2.4 \times 10^{-12}$ erg cm$^{-2}$ s$^{-1}$ in the 0.5--2.0 keV band.
We first selected 844 AGNs (i.e., sources with the {\tt Class} column = ``AGN'' in Table~2 of \citealt{Schwope}),
and then excluded 204 of them that are categorized as Seyfert 1.8--2 galaxies, narrow emission line galaxies, X-ray transient galaxies, BL Lac objects, or blazars, according to the {\tt Type} column in the \citet{Schwope} catalog to derive the reliable 2--10 keV fluxes (see Section \ref{Lx}) and Eddington ratios (see Section \ref{Edd_D}).
Finally, 640 type 1 AGNs with spectroscopic redshifts (available in the \citealt{Schwope} catalog) were selected for this work.

We compiled the ultraviolet (UV), optical, NIR, and MIR data of the selected type 1 AGNs, and cross-identified the sources in our sample with those in these multi-wavelength catalogs, adopting a search radius of 3$\arcsec$. 
The UV data were taken from the {\it Galaxy Evolution Explorer} \citep[{\it GALEX}:][]{Martin,Bianchi} satellite All-Sky Imaging Survey (AIS) data release (DR) 5, which provides the far-UV (FUV, 1344--1786 \AA) and near-UV (NUV, 1771--2831 \AA) photometries.
Before the cross-matching, we extracted 55,700,299 sources with {\tt Fexf} = 0 and {\tt Nexf} = 0, to ensure reliable FUV and NUV photometry, where {\tt Fexf} and {\tt Nexf} are {\tt SExtractor} \citep{Bertin} extraction flags for FUV and NUV, respectively (see user's manual\footnote{\url{https://www.astromatic.net/software/sextractor}} for more detail).
For the optical data, we used the ``forced photometry'' data\footnote{\url{https://outerspace.stsci.edu/display/PANSTARRS/PS1+ForcedMeanObject+table+fields}} of the $g$-, $r$-, $i$-, $z$-, and $y$ bands in the Pan-STARRS1 (PS1) survey \citep{Chambers,Flewelling,Magnier_16a,Magnier_16b,Magnier_16c,Waters} DR1\footnote{\url{https://outerspace.stsci.edu/display/PANSTARRS/PS1+Source+extraction+and+catalogs}}, which were obtained through the PS1 Catalog Archive Server Jobs System (CasJobs) service\footnote{\url{http://mastweb.stsci.edu/ps1casjobs/default.aspx}}.
As the NIR data, we adopted $J$-, $H$-, and $K_s$-band photometries in the Two Micron All Sky Survey \citep[2MASS;][]{Skrutskie_06} Point Source Catalog \citep[PSC;][]{Cutri_03} and Extended Source Catalog \citep[XSC;][]{Skrutskie_03}. 
We applied {\tt cc\_flg} = `000' for both NIR catalogs to extract objects without being affected by various artifacts (see Explanatory Supplement to the 2MASS All Sky Data Release and Extended Mission Product\footnote{\url{https://www.ipac.caltech.edu/2mass/releases/allsky/doc/explsup.html}}, for more detail). 
In addition, we utilized the {\it WISE} MIR data at 3.4, 4.6, 12, and 22 \micron, which are keys to derive the rest-frame 6 \micron \, luminosity of our AGN sample (see Section~\ref{L6}).
We extracted sources with ({\tt w1sat} = 0 and {\tt w1cc\_map} = 0) or ({\tt w2sat} = 0 and {\tt w2cc\_map} = 0) or ({\tt w3sat} = 0 and {\tt w3cc\_map} = 0), or ({\tt w4sat} = 0 and {\tt w4cc\_map} = 0) in the AllWISE catalog \citep{Cutri_14}, to have secure photometry at either band (see the Explanatory Supplement to the AllWISE Data Release Products\footnote{\url{http://wise2.ipac.caltech.edu/docs/release/allwise/expsup/index.html}}, for more detail).
We finally detected 323 (50.5 \%), 464 (72.5 \%), 575 (89.8 \%), and 579 (90.5 \%) counterparts among 640 AGNs in our sample, using the {\it GALEX} DR5, PS1 DR1, 2MASS PSC/XSC, and AllWISE catalogs, respectively.
We note that there are no multiple counterparts within 3$\arcsec$ in {\it GALEX} catalog.
However, 1/464 (0.2 \%) object has three PS1 counterparts while 8/464 (1.7 \%) objects have two PS1 counterparts.
Also, 1/575 (0.2 \%), and 1/579 (0.2 \%) objects have two candidates of counterparts in 2MASS, and ALLWISE, respectively.
In this study, we chose the closest object as the counterpart in each catalog.
The expected surface density of type 1 AGNs that are detectable by {\it GALEX}, PS1, 2MASS, and {\it WISE} is approximately 100 deg$^{-2}$  that is a rough estimate based on number counts \citep[e.g.,][]{DiPompeo,Ross_13}.
Thus, by adopting a search radius of 3$\arcsec$, the probability of chance coincidence is about 0.02 \% that is negligibly small.

We also checked the optical and IR images for 640 - 579 = 61 objects without {\it WISE} counterparts, and found that their optical coordinates in RBS catalog would be incorrect.
Therefore, we removed those 61 objects and focus on 579 AGNs in the following analysis.
Since the above issue (i.e., the refereed optical coordinates of some objects in RBS catalog might have a problem) is expected to be randomly occurred and their Eddington ratios are unavailable in \cite{Shen} (see Section \ref{Edd_D}), removing these objects less affects our correlation analysis.

\subsection{Hard X-ray luminosity}
\label{Lx}

We calculated the absorption-corrected, rest-frame 2--10 keV luminosity ($L_{\rm X}$) of each AGN in the following manner.
Even though only optical type-1 AGNs were treated in our study, a fraction of them may show absorption, typically of $N_{\rm H} <10^{22}$ cm$^{-2}$, in their X-ray spectra \citep[e.g.,][]{Ueda15}. 
Hence, for each source we checked its hardness ratio (HR) in the 0.1--2.0 keV band, defined as HR = $(H-S)/(H+S)$ where $H$ and $S$ are the vignetting-corrected 0.5--2.0 keV and 0.1--0.4 keV count rates, respectively. Using the energy response of the {\it ROSAT} Position Sensitive Proportional Counter, we first made conversion tables between HR and intrinsic absorption at the source frame ($N_{\rm H}$), for given redshift, photon index ($\Gamma$),  and Galactic absorption.
If the observed HR of a source was larger than that expected from an intrinsically-unabsorbed power law with assumed $\Gamma$ and the Galactic absorption, we estimated $N_{\rm H}$, by referring to the tables.
It is known that the photon index of an AGN increases with  the Eddington ratio \citep[e.g.,][]{Shemmer,Brightman,Trakhtenbrot}.
For 89 sources whose Eddington ratios were estimated in \cite{Shen}, we assumed a photon index calculated with the formula\footnote{We adopted $\Gamma= 0.167 \times \log \lambda_{\rm Edd} + 2.00.$} given by \citet{Trakhtenbrot}, whereas $\Gamma=1.9$ was assumed for the rest of the sample.
The distribution of $\Gamma$ for the 89 sources has a mean value of 1.87, which is close to 1.9, with a standard deviation of 0.09. We found that intrinsic absorption larger than  $10^{20}$ cm$^{-2}$ was required only in 29 sources, among which 25 showed $N_{\rm H} < 10^{21.5}$ cm$^{-2}$.
The absorption-corrected, rest-frame 2--10 keV luminosity was then calculated from the 0.5--2 keV flux listed in \citet{Schwope} (equivalent to the 0.5--2 keV  count rate), redshift, photon index, Galactic absorption, and intrinsic absorption (if any). 
For simplicity, we propagated only the statistical error in the 0.5--2 keV count rate into the luminosity error, because the uncertainty caused by that in HR (hence $N_{\rm H}$) was much smaller than it in most cases.
We finally multiplied the luminosity by 1.15 to correct for a possible cross-calibration difference in absolute fluxes between {\it ROSAT} and other major X-ray observatories (see \citealt{Ueda14}).
We have confirmed that this uncertainty little affects the following results.

\begin{figure}
\centering
\includegraphics[width=\hsize]{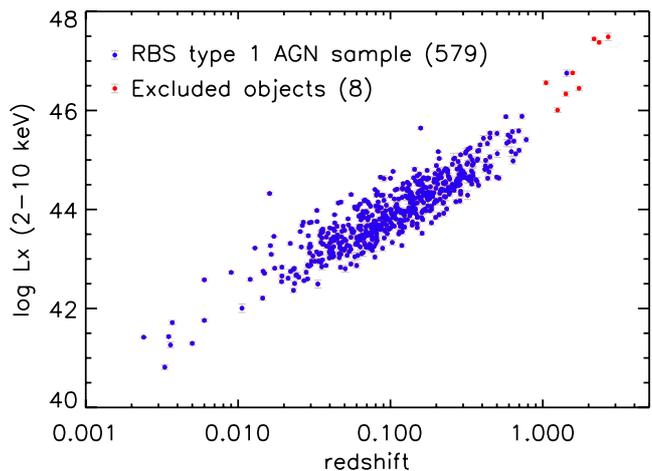}
\caption{The absorption-corrected, rest-frame 2--10 keV luminosity as a function of the redshift of the type 1 AGNs drawn from RBS catalog. Blue filled circles represent our sample of 579 type 1 AGNs, while red filled circles represent the candidates of blazars, which are not used in this work.
\label{z_Lx}}
\end{figure}

Figure \ref{z_Lx} shows the hard X-ray luminosity of the 579 type 1 AGNs, as a function of redshift.
An advantage of using the RBS catalog is that we can include 45/579 (7.8 \%) luminous quasars with $\log L_X$ $>$ 45. 
The X-ray luminosities of luminous quasars, especially the 9 sources with $\log L_X$ $>$ 46, however, may be amplified by the beaming effect, even if they are not classified as blazars in the RBS catalog.
Thus, we carefully checked the properties of those quasars, through the NASA/IPAC Extragalactic Database
 (NED\footnote{\url{https://ned.ipac.caltech.edu/}}) and the Set of Identifications,
Measurements, and Bibliography for Astronomical Data (SIMBAD\footnote{\url{http://simbad.u-strasbg.fr/simbad/}}) database, as well as the literature.
We found that 8/9 could be blazars, and removed them. 
Consequently, $579 -8 = 571$ type 1 AGNs with $0.002 < z < 1.436$ (mean $z$ = 0.15) and $40.81 < \log L_{\rm X} < 46.75$ (mean $\log L_{\rm X} = 43.9$) were left for further analysis.

\subsection{Mid-infrared luminosity}
\label{L6}

We estimated the rest-frame 6 \micron \, luminosity ($L_6$) of our type 1 AGN sample by performing the spectral energy distribution (SED) fitting.
One caution is that the contamination of the host galaxy emission to the NIR fluxes would affect the resultant $L_6$ value, especially for low-luminosity AGNs \citep[see e.g.,][]{Mateos}.
Therefore, we estimated how much the stellar components contribute to the NIR fluxes, through the SED fitting, and calculated the accurate $L_6$ values, contributed only by the AGN emission.

\begin{table}
\caption{Parameter ranges used in the SED fitting with {\tt CIGALE}.} 
\label{T1}
\centering                          
\begin{tabular}{l c }        
\hline\hline                 
Paramerer & Value\\          
\hline
\multicolumn2c{Delayed SFH}\\
\hline
$\tau_{\rm main}$ [Myr] & 500, 1000, 2000, 4000, 6000, 8000 \\
Age [Myr] & 500, 1000, 2000, 4000, 8000 \\
\hline
\multicolumn2c{SSP \citep{Bruzual}}\\
\hline
IMF & \cite{Chabrier} \\
Metallicity & 0.02 \\
\hline
\multicolumn2c{Dust attenuation \citep{Calzetti}}\\
\hline
$E(B-V)_*$ & 0.01, 0.2, 0.4, 0.6, 0.8, \\
& 1.0, 1.2, 1.4, 1.6, 1.8, 2.0 \\
\hline
\multicolumn2c{AGN emission \citep{Fritz}}\\
\hline
$R_{\rm max}/R_{\rm min}$ & 10, 30, 60, 100, 150 \\
$\tau_{\rm 9.7}$ & 0.1, 0.3, 0.6, 1.0, 2.0, 3.0, 6.0, 10.0 \\
$\beta$ &  $-1.00, -0.75, -0.50, -0.25, 0.00$ \\
$\gamma$ &   0.0, 2.0, 4.0, 6.0 \\
$\theta$ & 60., 100., 140 \\  
$\psi$ & 89.990 \\
$f_{\rm AGN}$ & 0.5, 0.6, 0.7, 0.8, 0.9, 0.99\\
\hline
\end{tabular}
\end{table}

\begin{figure*}
\includegraphics[width=\hsize]{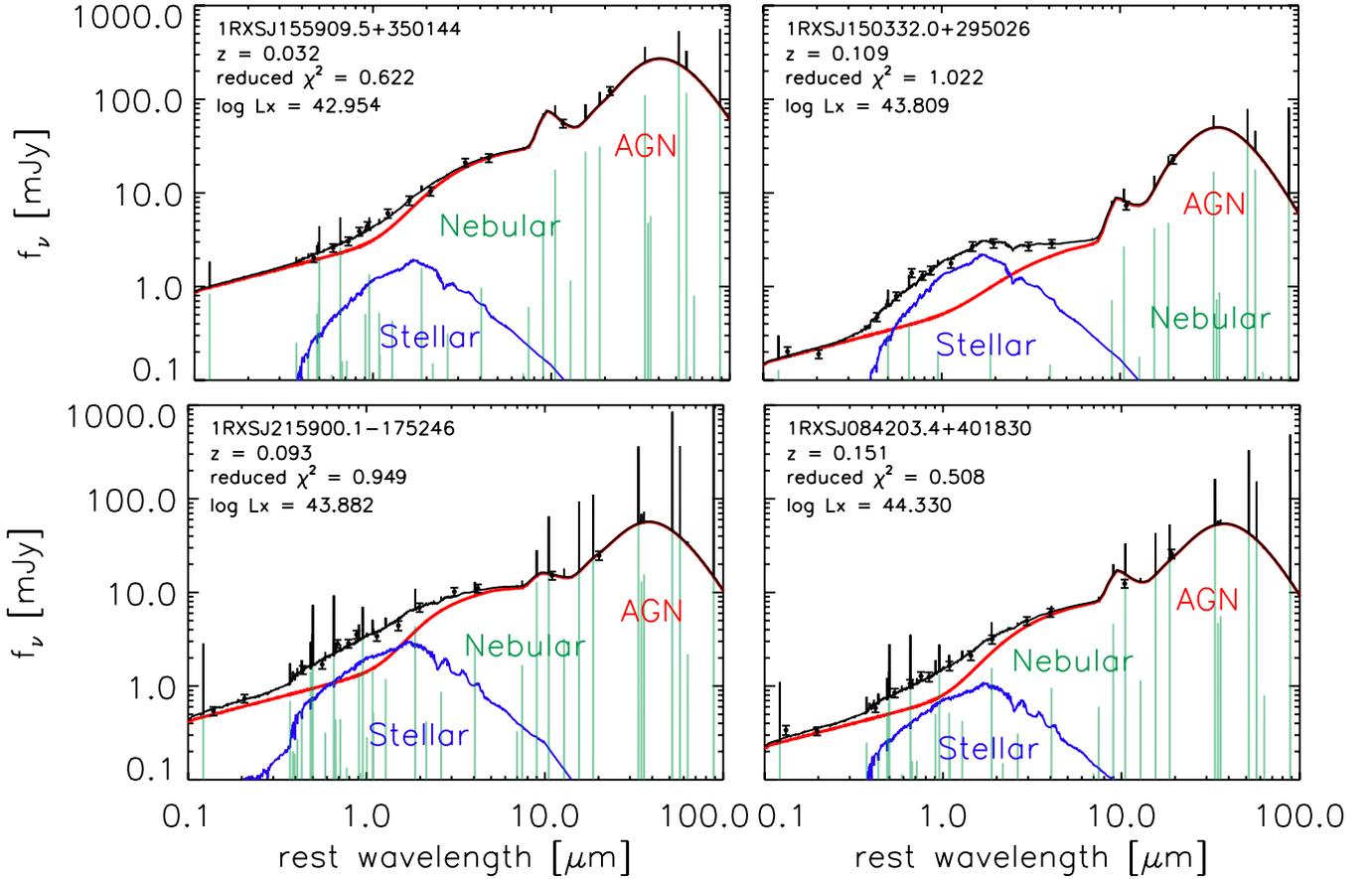}
\caption{Examples of the SED fitting to our sample with {\tt CIGALE}. The contribution of the stellar component is shown with blue lines while that of the AGN is indicated with red lines. Green lines represent nebular emission. The black solid lines represent the resultant SEDs (the sum of the stellar, nebular emission, and AGN). 
\label{SED}}
\end{figure*}

To perform a detailed SED modeling in a self-consistent framework, we employed {\tt CIGALE}\footnote{\url{https://cigale.lam.fr/2018/02/27/version-0-12-1/}} \citep[Code Investigating GALaxy Emission:][]{Burgarella,Noll,Boquien}, where we can handle many parameters such as the single stellar population (SSP), star formation history (SFH), attenuation law, AGN emission, and dust emission. 
We used a ``delayed'' SFH model, assuming a single starburst with an exponential decay.
We adopted the stellar templates provided from \cite{Bruzual} assuming the initial mass function (IMF) in \cite{Chabrier}, and the standard default nebular emission model included in {\tt CIGALE}. 
We also utilized models described by \cite{Calzetti} for the dust attenuation with a parameter of color excess of stellar component ($E(B-V)_{*}$).
For AGN emission, we used a model provided by \cite{Fritz}.
We note that only the type 1 AGN emission models were considered, by fixing the $\psi$ parameter (an angle between the AGN axis and the line of sight) at 89.990$\degr$ \citep[see][for more detail]{Fritz,Ciesla_15}. 
Table \ref{T1} lists the detailed parameter ranges adopted in the SED fitting.

Under the parameter setting described in Table \ref{T1}, we fitted the stellar and AGN models to at most 14 photometric points (FUV, NUV, $g$, $r$, $i$, $z$, $y$, $J$, $H$, $K_{\rm s}$-band, and 3.4, 4.6, 12, and 22 $\micron$ data) determined with {\it GALEX}, Pan-STARRS, 2MASS, and {\it WISE}.
For FUV and NUV data, we used ``flux\_nuv/fuv'' that is calibrated flux density through the pipeline of {\it GALEX} \citep[see][for more detail]{Martin}.
For optical data, we used ``g/r/i/z/yfapflux'' that is mean aperture flux from forced single epoch detections at each filter \cite[see][]{Flewelling}.
For NIR data, we basically used ``j/h/ks\_m'' that is a default magnitude at each band in 2MASS PSC for point sources while we used ``j/h/ks\_m\_ext'' that is magnitude from fit extrapolation in 2MASS XSC if they are available in that catalog (see Section 2.2.a and 4.5.e in Explanatory Supplement to the 2MASS All Sky Data Release and Extended Mission Products).
For MIR data, we estimated MIR flux densities in the same manner as \cite{Toba}; g1-4mag were used for objects with {\tt ext\_flg} $\neq$ 0 while w1-4mpro were used for objects with {\tt ext\_flg} = 0.
Since the photometry we utilized here is expected to trace the total flux at each band, we did not apply aperture correction.
We note that 505/571 ($\sim$88 \%) objects have at least 9 photometric data (249/518 objects have 14 photometric data) while only 5/571 ($\sim$0.9 \%) have only 4 photometric data, suggesting that SEDs of most objects in our sample are well determined. 

Figure \ref{SED} presents examples of the SED fitting of our sample with {\tt CIGALE}.
We confirmed that 428/571 ($\sim$75 \%) objects have reduced $\chi2 < 3.0$ while 488/571 ($\sim$85 \%) objects have reduced $\chi^2 < 5.0$.
Using the resultant SEDs, we estimated the rest-frame 6 \micron \, luminosities of pure AGNs.
We found that the contamination of the host galaxies was increased with decreasing X-ray luminosity,   typically about 7.2, 6.6, and 3.6 \% to the total 6 \micron \, luminosities for objects with $\log \,L_{\rm X} < 43$, $43 < \log \,L_{\rm X} < 44$, and $\log \,L_{\rm X} > 44$, respectively, which is in good agreement with what \cite{Mateos} reported.

\section{Result and discussion}

\begin{figure*}
\includegraphics[width=\hsize]{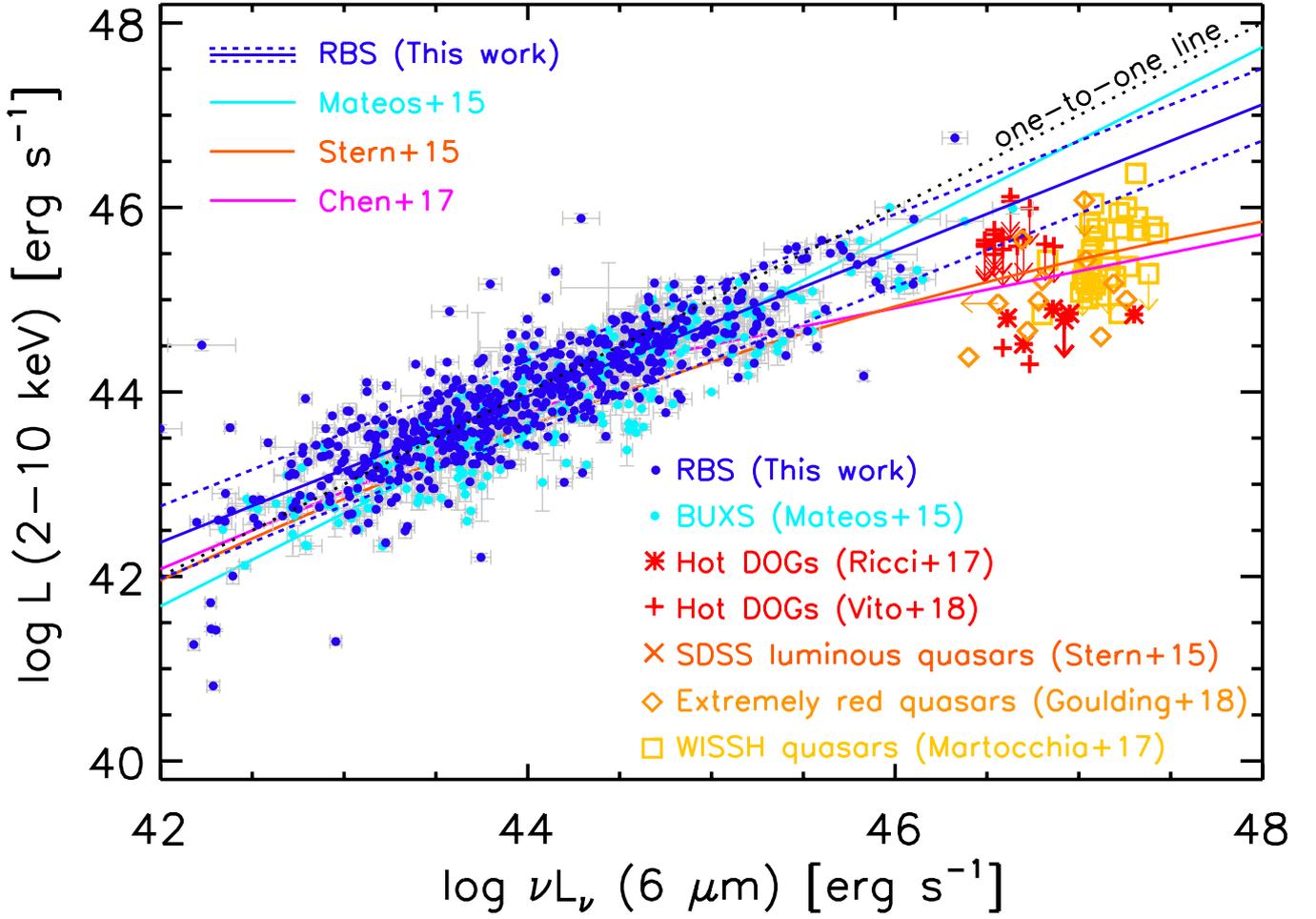}
\caption{Relation between the rest-frame 6 \micron \, luminosity contributed from AGNs and the absorption-corrected, rest-frame 2--10 keV luminosity, for various types of AGNs. Blue circles represent our type 1 AGN sample, selected from the RBS. The best-fit linear function in the log-log space and its 1$\sigma$ dispersions are shown with the blue solid line (see Equation \ref{Eq1}) and the dotted blue lines, respectively. Cyan circles represent an X-ray selected AGN sample in the BUS catalog \citep{Mateos}, and the cyan solid line represent its best-fit linear relation. Red asterisks and crosses show hot DOGs selected from {\it WISE} \citep{Ricci,Vito}. Orange crosses, diamonds, and yellow squares indicate the SDSS-selected luminous quasars \citep{Stern}, ERQs \citep{Goulding}, and WISSH quasars \citep{Martocchia}, respectively. The orange and magenta lines show the two-dimensional polynomial and bilinear relations from \citet{Stern} and \citet{Chen}, respectively. Black dotted line indicates the one-to-one correspondence between the 6 \micron \, and 2--10 keV luminosities.}
\label{L6X}
\end{figure*}

\subsection{The MIR and hard X-ray luminosity relation}
\label{L6LX}
Figure \ref{L6X} shows the resultant relation between the rest-frame 6 \micron \, luminosity and the rest-frame 2--10 keV luminosity, where the $L_{\rm 6}$ and $L_{\rm X}$ were corrected for the contamination of the host galaxy (see Section \ref{L6}) and the absorption (see Section \ref{Lx}), respectively.
We also plot the $L_{\rm 6}$--$L_{\rm X}$ relation for a hard X-ray selected sample consisting of 232 AGNs, drawn from the Bright Ultra-hard XMM-Newton Survey (BUXS) \citep{Mateos}, and those for the optical and MIR selected hyper-luminous AGNs: the hot DOGs \citep{Ricci,Vito}, SDSS-selected quasars \citep{Stern}, ERQs \citep{Goulding}, and WISSH quasars \citep{Martocchia}.

We then performed a linear regression in log-log space and a correlation analysis for our type 1 AGN sample by taking into account $L_{\rm 6}$ and $L_{\rm X}$ errors, in the same manner as \cite{Mateos}.
We first checked the Spearman rank correlation coefficients that is 0.86 with null hypothesis probability of $< 10^{-5}$, indicating a tight correlation with high significance level.
We also employed a Bayesian maximum likelihood method provided by \cite{Kelly} in which the regression assumes the following formula;

\begin{equation}
\log L_{\rm X} - \sigma_{\log L_{\rm x}} = A + B \times (\log L_{\rm 6} - \sigma_{\log L_{\rm 6}}) + \epsilon,
\end{equation}
where $A$ and $B$ are the regression coefficients (intercept and slope), and $\epsilon$ is the intrinsic random scatter about the regression.
$\sigma_{\log L_{\rm 6}}$ and $\sigma_{\log L_{\rm X}}$ are the 1$\sigma$ uncertainty of $L_{\rm 6}$ and $L_{\rm X}$, respectively.
We conducted a Markov chain Monte Carlo (MCMC) simulation where the Markov chain is created using the Metropolis-Hastings algorithm.
We assumed uniform prior distributions for the regression parameters, and estimated the linear correlation coefficient between $\log L_{\rm X} - \sigma_{\log L_{\rm x}}$ and $\log L_{\rm 6} - \sigma_{\log L_{\rm 6}}$.
The best-fit regression parameters and correlation coefficient with uncertainties are provided by calculating the mean and standard deviation from the posterior probability distributions of the model parameters with 10,000 iterations from the MCMC sampler \cite[see also][]{Mateos,Chen}.

The resultant linear relation is as follows;
\begin{equation}
\label{Eq1}
\log L_{\rm X} = (0.80 \pm 0.02) \log L_{\rm 6} + (9.14 \pm 0.93),
\end{equation}
and the correlation coefficient is $r = 0.84 \pm 0.01$, confirming again a tight correlation between $L_{6}$ and $L_{X}$ for out type 1 AGN sample.
This linear $L_{\rm 6}$--$L_{\rm X}$ correlation is roughly consistent with that reported by \cite{Mateos} and \cite{Chen} in the low luminosity region for $\log L_X < 44.5-45$.
We also performed a partial correlation analysis between $\log L_{\rm 6}$ and $\log L_{\rm X}$, with the correlation component due to redshift removed.
The resultant partial correlation coefficient is 0.61, which is still indicating a tight correlation.
We conclude that our type 1 AGN sample with {\it ROSAT} all-sky survey allows us to see a tight correlation for a wide luminosity range.
However, it cannot explain the $L_{\rm 6}$--$L_{\rm X}$ relations of hyper-luminous quasars selected from the SDSS/WISE (e.g., Hot DOGs, ERQs, and WISSH quasars); these objects show a negative offset with respect to the linear relation. To describe this non-linear behavior seen over a wide luminosity range, a luminosity-dependent relation would be required, as \cite{Stern} and \cite{Chen} suggested.

\subsection{Dependence of $L_{\rm X}$/$L_{\rm 6}$ on Eddington ratio}
\label{Edd_D}
In Section \ref{L6LX}, we confirmed that hyper-luminous quasars tend to be ``X-ray weak'' with respect to their MIR luminosities. 
Some possibilities were suggested in previous works, as the origin of this X-ray deficit in hyper-luminous quasars. 
For example, \cite{Chen} reported that the different $L_{\rm 6}$--$L_{\rm X}$ relations provided in the literature were due to the different X-ray flux limits. 
Other possibilities to explain the deficit may be a luminosity-dependent covering factor ($C_{\rm f}$) and may be a dependence of the MIR to X-ray luminosity ratio on hydrogen column density ($N_{\rm H}$).
However, \cite{Martocchia} concluded that the observed X-ray deficit in hyper-luminous quasars cannot be explained by a luminosity-dependent $C_{\rm f}$, and \cite{Mateos} found no clear dependence of the MIR to X-ray luminosity ratio on $N_{\rm H}$, in AGNs with $\log N_{\rm H} < 24$ cm$^{-2}$ \citep[see also][]{Lutz,Gandhi,Ichikawa_12}.

\begin{table*}
\caption{Summary of the resultant result for correlation analysis.}
\label{tbl1}
\centering
\begin{tabular}{lccc}
 \hline \hline
 Sample & N & $r$
 \\ \hline
RBS sample & 88 & $-0.60 \pm 0.07$ \\
RBS sample + SDSS quasars w/o RBS$^{a}$ & 232 & $-0.55 \pm 0.08$ \\
RBS sample + SDSS quasars w/o RBS$^{a}$ + BUXS sample$^{b}$ & 291 & $-0.34 \pm 0.07$ \\
RBS sample + SDSS quasars w/o RBS$^{a}$ + BUXS sample$^{b}$ + Hot DOG$^{c}$ + WISSH quasars$^{d}$ & 295 & $-0.41 \pm 0.07$ \\
\hline
\multicolumn{3}{l}{$^a$ \cite{Shen}}\\
\multicolumn{3}{l}{$^b$ \cite{Mateos}}\\
\multicolumn{3}{l}{$^c$ \cite{Ricci}}\\
\multicolumn{3}{l}{$^d$ \cite{Bischetti,Martocchia}}\\
\end{tabular}
\end{table*}

As described in Section \ref{intro}, the X-ray emission is considered to originate in Comptonization of the accretion disk emission in the hot corona, while the MIR emission is generated in the dust torus, which absorbs and reprocesses the UV/optical emission from the accretion disk. 
Thus, both the X-ray and MIR luminosities are associated with the disk emission, and the main driver of $L_{\rm X}$--$L_{\rm 6}$ correlation is also expected to be a physical quantity that is (more or less) related to the disk, which is responsible for UV/optical emission. 
Actually, \cite{Ricci} suggested $\alpha_{\rm OX}$ and $\kappa_{\rm 2-10 \,keV}$, as candidates of the main driver of that correlation.
Nevertheless, the above ideas, including the difference in the X-ray flux \citep{Chen}, are still based on observational facts.
In order to investigate what kinds of physics determine this X-ray deficit and $L_{\rm 6}$--$L_{\rm X}$ relation, we need to focus on a quantity that is directly associated with the physics of the accretion disk.
Here, we employ, as such a quantity, the Eddington ratio ($\lambda_{\rm Edd} = L/L_{\rm Edd}$), which is also expected to be an essential physical quantity to trace the AGN activity \citep[see also][]{Ricci,Goulding}.
Indeed, $\alpha_{\rm OX}$ and $\kappa_{\rm 2-10 \,keV}$ depend on $\lambda_{\rm Edd}$, although their correlations have large scatters \citep[e.g.,][]{Vasudevan,Lusso,Jin,Liu}.

We compiled the Eddington ratio given in \cite{Shen}, who provided physical properties (such as black hole mass and Eddington ratio) of the 105,783 quasars in the SDSS DR7 quasar catalog.
In that catalog, the black hole mass was estimated by using the H$\beta$ line for $z < 0.7$ and  Mg{\,\sc ii} for $0.7 < z < 1.9$, while the bolometric luminosity was computed from the monochromatic luminosities at 5100 \AA \,($z < 0.7$) and 3000 \AA \,($0.7 < z < 1.9$), by applying the spectral fitting and the bolometric corrections with correction factors of BC$_{5100}$ = 9.26 and BC$_{3000}$ = 5.15 \citep[see][and references therein]{Shen}. 
We found that the Eddington ratios of 88 type 1 AGNs in our sample were available in \cite{Shen}.
For 88 type 1 AGNs, we performed the correlation analysis between the Eddington ratios and luminosity ratio of the hard X-ray and MIR luminosities ($L_{\rm X}/L_{\rm 6}$), by using the procedure provided by \cite{Kelly} (see Section \ref{L6LX}).
The resultant correlation coefficient is $r = -0.60 \pm 0.07$ (see first line of Table \ref{tbl1}), indicating a tight correlation between $\lambda_{\rm Edd}$ and $L_{\rm X}/L_{\rm 6}$.

However, it should be noted that our type 1 AGN sample detected by {\it ROSAT} is biased toward brighter in X-ray flux.
In order to confirm the robustness of the correlation we found, we need to perform the correlation analysis for AGNs including X-ray fainter ones that are not detected by {\it ROSAT}.
We compiled SDSS quasars \citep{Shen} with counterpart of {\it WISE} but without counterpart of RBS where their $L_{\rm X}/L_{\rm 6}$ is treated as upper limit for our analysis that was estimated as follows.
We first estimated the Galactic $N_{\rm H}$ of each object\footnote{\url{http://www.swift.ac.uk/analysis/nhtot/}}.
We then calculated upper limit of flux at 2--10 keV for each object based on the detection limit of count rate in RBS (0.2 c s$^{-1}$) and obtained Galactic $N_{\rm H}$.
Since the redshift distribution between our sample and quasar sample in \cite{Shen} is different, we used only SDSS quasars with $z < 0.18$ whose mean $z$ is same as that of our sample.
In addition, we compiled BUXS type 1 AGN sample that is drawn from \cite{Mateos} where the detection limit of BUXS is about 100 times deeper than RBS.
We added SDSS quasars undetected by {\it ROSAT} and/or BUXS type 1 AGNs to our sample, and conducted correlation analysis in each case.
Since the method presented by \cite{Kelly} allows us to do a correlation analysis with taking into account not only x and y errors but also upper limit of either x or y, we again employed this code for our correlation analysis.

\begin{figure}
\includegraphics[width=\hsize]{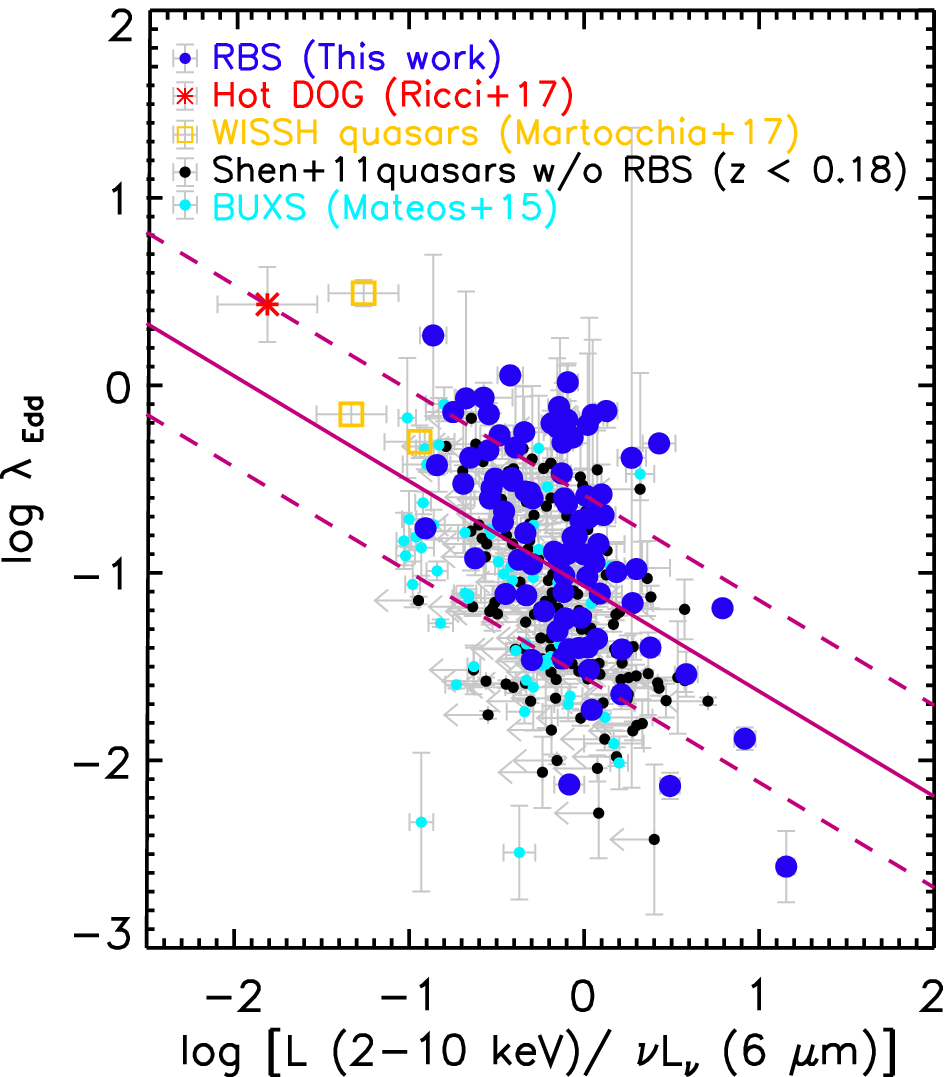}
\caption{Eddington ratio as a function of $L_{\rm X}/L_{\rm 6}$ of our RBS type 1 AGNs (blue circles), a hot DOG \citep[red asterisk:][]{Ricci}, WISSH quasars \citep[yellow squares:][]{Bischetti,Martocchia}, and BUXS type 1 AGNs \citep[cyan circles:][]{Mateos}. The black circles with left arrow represent SDSS quasars without RBS counterparts \citep{Shen}. The magenta solid line with dashed lines is the best-fit linear function with its 1$\sigma$ dispersions for RBS sample + the SDSS quasars without RBS counterparts, and the BUXS type 1 AGN sample (see Equation \ref{Eq2}).
\label{Edd}}
\end{figure}

The results of our correlation analysis are tabulated in Table \ref{tbl1}. 
Figure \ref{Edd} shows the Eddington ratios of our type 1 AGN sample, the SDSS quasars without RBS counterpart, and the BUXS AGNs as a function of the ratio between the hard X-ray and MIR luminosities ($L_{\rm X}/L_{\rm 6}$).
The best-fit linear function between $L_{\rm X}/L_{\rm 6}$ and $\lambda_{\rm Edd}$ derived through the regression analysis \citep{Kelly} is
\begin{equation}
\label{Eq2}
\log \lambda_{\rm Edd} = -(0.56 \pm 0.10) \log \, \left(\frac{L_{\rm X}}{L_{\rm 6}} \right) - (1.07 \pm 0.05).
\end{equation}
The correlation coefficient is $r = -0.34 \pm 0.07$ indicating a moderately tight correlation even taking into account AGNs that are undetected by {\it ROSAT} as well as those detected by {\it XMM-Newton}.
On the other hand, we found that adding X-ray fainter objects to our AGN sample make correlation weak, which means that a tight correlation ($r = -0.60$, see Table \ref{tbl1}) for our sample is due to the shallow flux limit of RBS.

We also assembled Eddington ratios for a hot DOG \citep{Ricci} and WISSH quasars \citep{Bischetti,Martocchia} where if uncertainty of $\lambda_{\rm Edd}$, $L_{6}$, or $L_{X}$ was not provided, we conservatively assumed 20 \% error of the corresponding quantity
We confirmed that our conclusion is not significantly affected if we assumed more larger or smaller error.
Figure \ref{Edd} also shows  the Eddington ratios of the hot DOG and  the WISSH quasars as a function of $L_{\rm X}/L_{\rm 6}$.
We confirmed that those objects showing X-ray deficit (see Figure \ref{L6X}) tend to have large $\lambda_{\rm Edd}$.
Eventually, we obtained an evidence that the Eddington ratio decreases with increasing $L_{\rm X}/L_{\rm 6}$, indicating that AGNs with high Eddington ratios tend to be X-ray weak with respect to the MIR emission.

Why do AGNs with high Eddington ratios tend to be X-ray weak?\
One possibility is a change in the geometry of the accretion flow, like what is suggested in Galactic black hole binaries \citep[e.g.,][]{Esin,Done}; at low $\lambda_{\rm Edd}$, the optically-thick, geometrically-thin accretion disk is thought to be truncated at some radii and replaced by a hot inner flow (or ``corona''), producing the Comptonized component in the hard X-ray band. 
The inner edge of the disk moves inwards as $\lambda_{\rm Edd}$ increases, and accordingly the size of the hot corona becomes smaller. 
Consequently, the solid angle of the corona viewed from the disk is reduced, leading a lower probability of inverse Compton scattering.
This effect could explain the observed X-ray deficit of AGNs at high Eddington ratios. 
Similar effects were inferred in previous studies of AGNs \citep[e.g.,][]{Kubota}.

The MIR luminosity is naively described as $L_{\rm MIR} = C_{\rm f} L_{\rm bol}$, where $C_{\rm f}$ is the covering factor of the dust torus. \cite{Ricci_17b} reported that $C_{\rm f}$ is almost constant ($\sim 0.3$) for AGNs with $\log \lambda_{\rm Edd} > -1.5$, while it rapidly increases to 0.8, once $\log \lambda_{\rm Edd}$ becomes lower than $-1.5$.
Because most of the AGNs in Figure \ref{Edd} have $\log \lambda_{\rm Edd} > -1.5$ and thus $C_{\rm f}$ is considered as constant, $(L_{\rm X}/L_{\rm 6})$ is basically proportional to $\kappa^{-1}_{\rm 2-10 \,keV}$.
\cite{Lusso} found that $\kappa_{\rm 2-10\, keV}$ correlates with $\lambda_{\rm Edd}$, as $\log \, \kappa_{\rm 2-10 \, keV} /\log \, \lambda_{\rm Edd} \sim 1.2$.
If we adopt this value, we obtain $\log \, \lambda_{\rm Edd}/ \log \,(L_{\rm X}/L_{\rm 6}) \sim -0.8$, which is roughly consistent with what we obtained from the fitting for all data points in Figure \ref{Edd}.
We note that some AGNs with $\log \lambda_{\rm Edd} < -1.5$ seems to be deviated from the best-fit relation in Figure \ref{Edd}.
This is interpreted as an increase of $C_{\rm f}$, which boosts their 6 \micron \, luminosity and thereby makes their $L_{\rm X}/L_{\rm 6}$ values smaller than that expected from the best-fit function.
In summary, the dependence of $L_{\rm X}/L_{\rm 6}$ on $\lambda_{\rm Edd}$ can be understood as a change in the structure of the accretion flow.

\section*{Acknowledgements}
We gratefully acknowledge the anonymous referee for a careful reading of the manuscript and very helpful comments.
We are deeply thankful to Dr. Denis Burgarella, Dr. M\'ed\'eric Boquien, and Dr. Laure Ciesla for helping us to understand {\tt CIGALE} code.
This work is based on archival data from the {\it Galaxy Evolution Explorer} which is operated for NASA by the California Institute of Technology under NASA contract NAS5-98034.
The Pan-STARRS1 Surveys (PS1) and the PS1 public science archive have been made possible through contributions by the Institute for Astronomy, the University of Hawaii, the Pan-STARRS Project Office, the Max-Planck Society and its participating institutes, the Max Planck Institute for Astronomy, Heidelberg and the Max Planck Institute for Extraterrestrial Physics, Garching, The Johns Hopkins University, Durham University, the University of Edinburgh, the Queen's University Belfast, the Harvard-Smithsonian Center for Astrophysics, the Las Cumbres Observatory Global Telescope Network Incorporated, the National Central University of Taiwan, the Space Telescope Science Institute, the National Aeronautics and Space Administration under Grant No. NNX08AR22G issued through the Planetary Science Division of the NASA Science Mission Directorate, the National Science Foundation Grant No. AST-1238877, the University of Maryland, Eotvos Lorand University (ELTE), the Los Alamos National Laboratory, and the Gordon and Betty Moore Foundation.
This publication makes use of data products from the Two Micron All Sky Survey, which is a joint project of the University of Massachusetts and the Infrared Processing and Analysis Center/California Institute of Technology, funded by the National Aeronautics and Space Administration and the National Science Foundation.
This publication makes use of data products from the {\it Wide-field Infrared Survey Explorer}, which is a joint project of the University of California, Los Angeles, and the Jet Propulsion Laboratory/California Institute of Technology, funded by the National Aeronautics and Space Administration.
This research has made use of the NASA/ IPAC Infrared Science Archive, which is operated by the Jet Propulsion Laboratory, California Institute of Technology, under contract with the National Aeronautics and Space Administration.
This research has made use of the NASA/IPAC Extragalactic Database (NED), which is operated by the Jet Propulsion Laboratory, California Institute of Technology, under contract with the National Aeronautics and Space Administration.
This research has made use of the SIMBAD database, operated at CDS, Strasbourg, France. 
Y.Toba, W.H.Wang, and Y.Y.Chang acknowledge the support from the Ministry of Science and Technology of Taiwan (MOST 105-2112-M-001-029-MY3).
This work is supported by JSPS KAKENHI Grant numbers 18J01050 (Y.Toba), 17K05384 (Y.Ueda), 16K17672 (M.Shidatsu), 16H01101, 16H03958, and 17H01114 (T.Nagao), 15H02070, and 16K05296 (Y.Terashima).
K.Matsuoka is supported by JSPS Overseas Research Fellowships.






\bsp	
\label{lastpage}
\end{document}